\documentclass[twocolumn,showpacs,showkeys,preprintnumbers,superscriptaddress,amsmath,floatfix,amssymb,secnumarabic]{revtex4}
\usepackage[colorlinks=true]{hyperref}
\usepackage{graphicx}
\usepackage{epsfig}

\newcommand{\beq}{\begin{equation}}
\newcommand{\eeq}{\end{equation}}

\def\dalam{\hbox
{\vrule\vbox{\hrule\hbox to 1ex{ \hfill}\kern 1 ex\hrule}\vrule}}
\def\sign{\hbox{sign}}
\def\1/2{\hbox{$ {1 \over 2}$ }}

\def\h{\hbar}
\def\i/h{{i \over \h}}

\def\th{\tanh}

\def\inf{\infty}
\def\pd{\partial}

\def\a{\alpha}

\def\g{\gamma} \def\tg{\tilde {\g}}\def\G{\Gamma} 
\def\d{\delta} \def\D{\Delta}
\def\l{\lambda} 
 
 \def\tE{\tilde {E}}

\def\s{\sigma}\def\S{\Sigma}
 
\def\x{\xi}

 \def\F{\Phi}
 
\def\p{\psi}

\def\k{\kappa} \def\tk{\tilde \kappa}

\def\W{\Omega}
\def\tt{\theta}

\def\<{\langle}
\def\>{\rangle}

\def\({\left(}
\def\[{\left[}
\def\){\right)}
\def\]{\right]}

\begin{document}
\sloppy

\title{Quantum confinement under Neumann condition: atomic H filled in a lattice of  cavities}

\author{K.~Sveshnikov}
\email{costa@bog.msu.ru} \affiliation{Department of Physics and
Institute of Theoretical Problems of MicroWorld, Moscow State
University, 119991, Leninsky Gory, Moscow, Russia}

\author{A.~Roenko}
\email{roenko@physics.msu.ru} \affiliation{Department of Physics and
Institute of Theoretical Problems of MicroWorld, Moscow State
University, 119991, Leninsky Gory, Moscow, Russia}

\date{February 4, 2013}

\begin{abstract}
Energy spectrums of a  nonrelativistic particle and an H-like
atom  in a spherical box of size $R$ with general conditions of ``not
going out'' through the box surface are explored. The lowest energy levels reconstruction is described from the point of view of their asymptotical behavior for large $R$. The role of  von
Neumann-Wigner level reflection/avoided crossing effect in this spectrum
reconstruction is emphasized. The properties of  atomic H ground state in  a cell, formed by a spherical cavity with an outer potential shell and Neumann  condition on the outward boundary, are studied in detail. Some of them turn out to be quite new. The relevance of such a cell to a cubic lattice of cavities, occupied by H, is discussed be means of first principles and assumptions of the Wigner-Seitz model.
\end{abstract}


\maketitle

\subsection*{1. Introduction}

Considerable amount of theoretical and experimental activity has
been focused recently on spatially confined atoms and molecules
\cite{Jask}-\cite{Maris}.  So far, starting from the works of Michels \cite{Michels} and  Sommerfeld \cite{Sommerfeld}, main attention has been devoted to the properties of atoms and molecules, confined by an impenetrable or partially
penetrable potential barrier (\cite{Aquino}, \cite{Sen1} and refs. therein).  However,   in reality  general boundary conditions of ``not going out''  don't unavoidably imply  genuine trapping of a particle by
a cavity, rather they could in some special cases correspond to a
quite different picture, where the particle state undergoes
delocalization from the box with definite symmetry properties of
the wavefunction, as in the Wigner-Seitz model of alcaline metal
\cite{WS}. The latter circumstance turns out to be quite important,
since in some cases the cavities, where a
particle or an atom could reside, form a lattice, similar to that of an alcaline metal,  like
certain interstitial sites of a metal supercell, e.g. octahedral
positions of palladium fcc lattice \cite{Alefeld}-\cite{PdHx}. In this case a particle  (or valence atomic electron, provided that the whole lattice of cavities is occupied by atoms) finds itself in a periodic potential of a cubic lattice, and so the description of its ground state could be based on the first
principles of the Wigner-Seitz model \cite{WS}. With the same assumptions as  in \cite{WS}, it turns out to be a special type of ``confinement''  under Neumann boundary condition in the corresponding Wigner-Seitz cell.

The purpose of this letter is to explore the features of such a type of ``confinement'' state in a  cell, formed by a spherical cavity of radius $R$ with an outer potential shell of physically
reasonable width and depth, and Neumann  condition on the outward boundary. A number of nontrivial properties of such state, a part of which being similar to those described earlier for atoms trapped endohedrally inside a fullerene molecule \cite{Connerade1}, and more recently by means of general reflecting  boundaries \cite{Wiese}, while another part being quite new, is discovered by studying the asymptotical behavior of energy levels for large $R$.  Moreover, such an approach  allows for a valuable analysis of conditions, under which such phenomena  could take place. In particular, we describe the case, when the ground state of atomic H considered as a function of $R$, contains a  deep and strongly
pronounced well, where the bound energy could be remarkably larger than
that of  $1s$-level of the free atom $E_{1s}$, as well as the situation, when the
lowest level reveals  slowly decreasing power asymptotics for large $R$ and so its bound energy could exceed   $E_{1s}$  for actual nanocavities with $R \sim 100-1000$ nm.

\subsection*{2. General treatment  of a ``not going out'' state}

Stationary  state of a particle with mass $m$ confined in a vacuum
cavity $\W$ with boundary $\S$ should be described by an energy functional of
the following form \beq \label{f1} E[\p]=\int_\W \! d \vec r \
\left[ {\h^2 \over 2m } | {\vec \nabla } \p|^2 + U(\vec r ) \
|\p|^2 \right] +
\nonumber \\ \eeq
\beq \label{f1}
 + \ {\h^2 \over 2m} \int_\S \! d\s \ \l (\vec r) \
|\p|^2  \ , \eeq
 where $U(\vec r)$ is the potential   inside  $\W$, while the surface term $\int_\S$ corresponds to  contact interaction of the particle with medium, in which the cavity has been  formed, on the cavity boundary. The properties of this surface interaction are given by a real-valued function $\l (\vec r)$.

From the variational principle with normalization condition  $\<
\p|\p \>=\int_\W \! d \vec r \  |\p|^2 =1$ it follows that \beq
\label{f2}
 \left[ -{\h^2 \over 2m } \D + U(\vec r )  \right] \p=E \p
\eeq inside $\W$ combined with boundary condition imposed on $\p$
on the surface  $\S$ \beq \label{f3} \left. \[ \vec n \vec \nabla +
\l(\vec r )  \] \p \right|_\S=0 \ , \eeq with $\vec n $ being the
outward normal to $\S$.

Boundary condition (\ref{f3}) is known in mathematical physics  as
Robin's (or third kind) condition, under which the spectral
problem (\ref{f2}-\ref{f3}) is self-adjoint and so contains all the
required properties for a correct quantum-mechanical description of
a nonrelativistic particle confined in $\W$ \cite{Wiese},\cite{Sen2}.
The particle ``not going out''  property is fulfilled
here via vanishing normal to $\S$ component of the  quantum-mechanical flux
\beq \label{f4} \left.  \vec n \vec j \right|_\S=0
\ , \eeq where \beq \label{f5} \vec j = {\h \over 2m i} \ \(
\p^{\ast} \vec \nabla  \p - \p \vec \nabla \p^{\ast} \) \ . \eeq
At the same time, tangential components of  $\vec j$ could be
remarkably different from zero on $\S$ and so the particle could
be found quite close to the boundary with a marked probability. In
particular, such a picture takes place  in the Thomas-Fermi model
of many-electron atom \cite{LL}, as well as in  quark bag models of hadron
physics \cite{MIT bag},\cite{Chiral bag}.

When $\l=0$, the interaction of the particle with environment is
absent and so eq. (\ref{f3}) transforms into  Neumann (second kind)
condition \beq \label{f6} \left.  \vec n \vec \nabla \p
\right|_\S=0 \ , \eeq what corresponds to the boundary condition
of confinement for a scalar field in relativistic bag models \cite{MIT bag}.
Moreover, condition (\ref{f6}) appears in the Wigner-Seitz model of
an alcaline metal \cite{WS} and describes delocalization of valence
electrons creating the metallic bond, by continuing the atomic
wavefunction periodically  in the lattice. Indeed such a ``confinement'' state is  at the aim of our study.

If $\l \to \infty$, then   (\ref{f3}) turns into the Dirichlet
condition \beq \label{f7} \left.  \p \right|_\S=0 \ , \eeq and so
describes  confinement by an impenetrable barrier.

There are two well-established and quite important inequalities
for the ground state energy in the Dirichlet and Neumann cases of
confinement \cite{Sen2}. The first one takes place for the Dirichlet
problem (\ref{f7}) and tells, that if the volume $\W$ is embedded in
volume $\W_1$, then $E(\W) > E(\W_1)$ for any nonsingular $U(\vec r)$. Another
one concerns the Neumann case (\ref{f6}) and gives the following
estimate for the ground state energy
\beq \label{f8} E_{ground}
(\W) < { \int_\W  d \vec r \  U(\vec r ) \over \int_\W  d \vec r }
\  . \eeq
The inequality (\ref{f8}) follows immediately from
the variational principle, if one considers a constant trial
wavefunction in order to  fulfill the boundary condition (\ref{f6}) in
the simplest way. It can be easily generalized to the Robin's case
(\ref{f3}) in the following fashion. Let us consider the
confinement state in $\W$  with no surface interaction, but with
modified potential function \beq \label{f9} U_1(\vec r)=U(\vec r) +
{\h^2 \over 2m} \ \l (\vec r) \ \d_{\S_1}(\vec r) \ , \eeq where
$\d_{\S_1}(\vec r)$ denotes  surface $\d$-function with  $\S_1$
being a surface embedded in $\W$. The additional term in the modified
potential $U_1$ gives rise to the following contribution to the energy
functional \beq \label{f10} \D E[\p]= {\h^2 \over 2m} \int_{\S_1}
d\s \ \l (\vec r) \ |\p|^2  \ , \eeq but at the same time doesn't
affect the Neumann boundary condition on $\S$. The latter makes it
possible to draw a direct analogy with the procedure leading to
the inequality (\ref{f8}), since the trial wavefunction can be still
chosen as a constant throughout  $\W$. If we consider now the
limit  $\S_1 \to \S$ from the inside of  $\W$, then the region
$\W_1 \subset \W $ surrounded by the surface $\S_1$ tends to $\W$,
while the contribution of the region $\W-\W_1$ to energy becomes
negligibly small. Proceeding  further this way, we get the
expression (\ref{f1}) as the limiting point for the energy
functional, and so the following inequality for the ground state
energy \beq \label{f11} E_{ground} (\W) < { \int_\W  d \vec r \
U(\vec r ) \over \int_\W  d \vec r } \ + {\h^2 \over 2m} {\int_\S
d\s \ \l (\vec r) \over \int_\W  d \vec r } \  . \eeq The estimates
(\ref{f8},\ref{f11})  turn out to be quite effective for
understanding the ground state properties, especially for the case
of extremely small cavities.

\subsection*{3. Robin's reflecting boundaries}

Now let us consider the case of Robin's boundary condition (\ref{f3}). Since it has been already studied in \cite{Wiese},\cite{Sen2},\cite{Pupyshev},\cite{PEPAN}, we'll
point out here only those details, which are required for dealing
with a  more complicated and realistic model described in the next
section.

First example is a particle in a spherical potential well of
radius $R$ with a constant potential $U(\vec r)=U_0 \ , \ r<R \ ,$
and surface interaction $\l=$Const \cite{Wiese},\cite{PEPAN}.  In what follows, in order to
provide an effective comparison of  results, obtained for quite
different systems, we'll use relativistic units $\h=c=1$,
wavenumber and energy will be expressed in units of the particle mass
$m$, while distances --- in units of the particle Compton length
$1/m$. Considering $U_0$ as a reference point for the particle
energy, for $s$-levels  one obtains
\beq \label{f12} \tan kR = { kR
\over 1- \l R } \ , \eeq
where $k=\sqrt{2E}$.

It is easy to see from (\ref{f12}), that the energy levels considered
as functions of $R$ reveal remarkably different behavior
depending on the sign of $\l$. More concretely, when $\l >0$ and so  describes reflection between
the particle and environment, for  $R\to 0$ the wavenumber of the lowest
energy level behaves like $\sqrt{3\l/R}$, while the ground state
energy increases in the following way
\beq \label{f13} E_{ground}
(R) \to  {3 \l \over 2 R} \ , \ R \to 0 \ , \eeq
what follows directly from eq. (\ref{f12}) as well as from  the estimate (\ref{f11}). Such behavior of $E_{ground}(R)$ confirms, that for confined systems the standard uncertainty relation  should be replaced by a generalized one, which doesn't imply, that for
$R\to 0$ the kinetic energy of the particle could be
estimated as $O(1/R^2)$ (see \cite{Wiese},\cite{Wiese1} and discussion therein). The latter should be definitely correct in the case  $\l \to \infty$ only, i.e. in the case of genuine
trapping of the particle in a cavity by an impenetrable potential
wall. For eq.(\ref{f12}) such behavior occurs for the particle
states with positive energy  in the case of surface attraction $\l
\leq 0$, when for  $R\to 0$ the wavenumber of the lowest positive
level behaves like  $C/R$ with $C=4.49341$ being the first root of
the equation $\tan x =x$, while the  energy --- like $C^2/2R^2$.
For $R \to \infty$ both types of solutions for the lowest positive
level reveal the same asymptotics $E(R) \to \pi^2/2R^2$, what
corresponds to the Dirichlet condition (\ref{f7}).

For  $\l < 0$, i.e. for the case of
attraction between the particle and environment, the  generalized uncertainty relation for confined systems \cite{Wiese},\cite{Wiese1} provides, that  the ground state
$s$-level lyes  below the well's bottom and so should be found
from eq.(\ref{f12}) via   $k \to i\k$, i.e. from equation
\beq
\label{f14} \tanh \k R = { \k R \over 1+ |\l| R } \ . \eeq For $R
\to 0$
the wavenumber $\k(R)$ reveals the asymptotics
$\sqrt{3|\l|/R}$,   hence
\beq \label{f15} E_{ground} (R) \to  -{3
|\l| \over 2 R} \ , \ R \to 0 \ , \eeq
what could be easily
verified by  estimate (\ref{f11}) again.  There are no
contradictions with the general properties of the energy spectrum of
a nonrelativistic particle here, since for $\l < 0$ the surface term
in the expression (\ref{f1})  could be arbitrarily negative due to
$|\p|^2$ on the box boundary, which   might be  now arbitrarily
large without violating the normalization condition.

It should be  noted also,  that for $R \to \infty$ such a level
reveals the following asymptotics
\beq \label{f16}
 E_{ground} (R) \to -\l^2/2 - |\l|/R + O(1/R^2) \ , \ R \to \infty \ ,
\eeq
and so its behavior on the whole half-axis   $0 \leq R\leq
\inf$ should be quite similar to a shifted downwards hyperbole.

Therefore  for $\l <0$ the particle lowest  $s$-level lyes below the well's
bottom even in the case of increasing well's radius, but this
property cannot be detected from estimate (\ref{f11}). The latter
circumstance should be quite evident, since for large $R$ the
constant trial wavefunction cannot be a good approximation to the
genuine wavefunction of the problem.

This example shows explicitly, that  the spectrum of stationary
states of a particle confined in a box with  general ``not going
out'' conditions could reveal features, which are quite different
from  the deconfinement case. In
particular, for $\l <0$ the behavior of the ground
state  is such, that  the  energetically most favorable
state of a particle is to be caught by the smallest cavity.

The ``not going out'' state of  atomic H with nuclei charge
$q$ in a spherical cavity with radius $R$ and boundary conditions (\ref{f3}) turns out to be even
more specific \cite{Wiese},\cite{Sen2},\cite{Pupyshev},\cite{PEPAN}.  As in the previous case, surface interaction is given by
the constant $\l$, while motionless point-like atomic nuclei is  in
the center of the cavity, then  spherical symmetry is maintained
and the ground state energy minimized. From the solution of the
 Schroedinger-Coulomb problem   for the  radial wavefunction of the
electron state with orbital momentum $l$  one obtains up  to a
numerical factor \cite{LL}
\beq \label{f17} R_l (r) =e^{-\g r} r^l \ \F (b_l , c_l , 2 \g r) \ , \eeq
where
\beq \label{f18} \g=\sqrt{-2E} \ ,
\ b_l=l+1-q/\g \ , \ c_l=2l+2 \ , \eeq and  $\F(b,c,z)$ is the
confluent hypergeometric function of the first kind (Kummer
function). Definition, notations and main properties of the Kummer
function follow ref. \cite{Beytmen}. Substituting (\ref{f17}) into the boundary
condition  (\ref{f3}) yields the following equation for  energy
levels
\beq \label{f19}
\[ q/\g + (\l-\g)R -1 \]  \F_R + \[l+1 -q/\g\] \F_R(b+) =0 \ ,
\eeq
where
\beq \label{f20} \F_R=\F (b_l , c_l , 2 \g R) \ , \
\F_R(b+)=\F (b_l+1 , c_l , 2 \g R) \ . \eeq

As in the previous case of a  potential well, the most significant
changes in the spectrum take place for $R \to 0$, what  could be
seen at once from the  estimate (\ref{f11}). Here it should be noted, that for  atomic H the limit $R \to 0$ takes some care, since  relativistic effects give rise to the restriction $R \geq 10$ for the cavity sizes, where such an approach to the confinement problem, based on boundary condition (\ref{f3}), should be valid \cite{PEPAN}. So in what follows the limit $R \to 0$ should be understood either as a purely mathematical property of equations under consideration, or as decreasing $R$ up to $R \sim 10$. To underline the existence of this lower limit, the curves shown on Figs. 2-6 below will start from $R = 10$ too.

There are two  types of the
lowest level of atomic H in dependence on
relation between $\l$ and $q$. The first one takes place under assumption, that for
$R \to 0$ the wavenumber $\g$ remains finite, and so in the
vicinity of $R=0$ it could be represented by a series
\beq \label{f21} \g(R)=\g_0+ \g_1 R + \g_2 R^2 + ... \ . \eeq Expanding
$\F_R \ , \F_R(b+)$
in a power series in  $R$ (what is always
possible, since the Kummer series converges everywhere in the
complex plane), to the lowest order one obtains from (\ref{f19})
that $l=0$, and by proceeding further
\beq \label{f22} \l=q \ ,
\quad \g_0^2=q^2 \ , \quad \g_n=0, \quad  n \geq 1  \ . \eeq
It follows from (\ref{f22}), that if $\l=q$, then the ground state
energy of  atomic H  in a cavity for any $0 \leq R \leq \inf $
precisely coincides with that of $1s$-level of the free atom \beq
\label{f23} E_{ground}(R) = E_{1s}=-q^2/2 \  , \eeq what has been
already  mentioned in \cite{Sen2},\cite{Pupyshev}.

More precisely, for $l=0, \ \l=q, \  \g_0=\pm q$ eq. (\ref{f19}) is
satisfied for all $R$. For $\g_0=q$ it is provided by $b_0=0$ and
$\F(0,2,z)=1$, while for $\g_0=-q$ one obtains  $b_0=2 \ , \
\F(2,2,z)=e^z \ , \ \F(3,2,z)=(z/2+1)e^z$, and in both cases
substitution  into (\ref{f19}) gives an identity. There is however no
twofold  degeneracy of the level, since both signs in $\g_0=\pm
q$ correspond to the same radial $1s$-function $R_0(r)=A e^{-qr}$,
what should be quite obvious, because the parameter $\g$ is
defined via relation  $E=-\g^2/2$, where  the sign of $\g$ isn't
fixed.

As for a particle in a potential well, another type of levels
reveals for $R \to 0$ asymptotic behavior  similar to (\ref{f13})
or (\ref{f15}) and is found by assumption, that in the vicinity of
$R=0$ the wavenumber $\g$ is represented by a series \beq
\label{f24} \g(R)= {\x \over \sqrt{R}} + \x_0 + \x_1 \sqrt{R} +
\ldots \  . \eeq Substituting (\ref{f24}) into eq.(\ref{f19}), to the
lowest order in  $\sqrt{R}$ one obtains again $l=0$, while  higher
orders of expansion in   $\sqrt{R}$ yield \beq \label{f25}
\x^2=3(q-\l) \ , \quad \x_0=0 \ , \quad \x_1={q^2 + 3q\l +6\l^2
\over 20\x} \ , \quad \ldots \ . \eeq

As  a result, for such  type of  $s$-levels of  H in a
cavity one obtains the following dependence on the cavity radius for
$R \to 0$
\beq \label{f26} E_{ground}(R) \to - { 3(q-\l) \over 2 R}
- {q^2 + 3q\l +6\l^2  \over 20} + O(\sqrt{R}) \ ,
\nonumber \\ \eeq
\beq \label{f26} R  \to 0  \ . \eeq
Qualitative explanation of linear dependence on $q$
and  $\l$ is quite simple. As for a particle in a spherical well,
for $R \to 0$ the atomic wavefunction of such  $1s$-level inside a cavity
becomes almost constant, and so the estimate (\ref{f11}), which
reproduces the first term in (\ref{f26}), turns out to be  almost
exact too. The numerical solution of eq.(\ref{f19}) for   $q=\a \simeq 1/137$
and $\l=(1\pm 0.01) q \ $, $\ \l=(1\pm 0.02) q$  shows, that  the
behavior of such  $s$-levels tends to the asymptotics (\ref{f26})
for $R$ of order  about several tenths of $a_B=1/\a \simeq 137$ (Fig.1).

\begin{figure}
\includegraphics[width=8.6 cm]{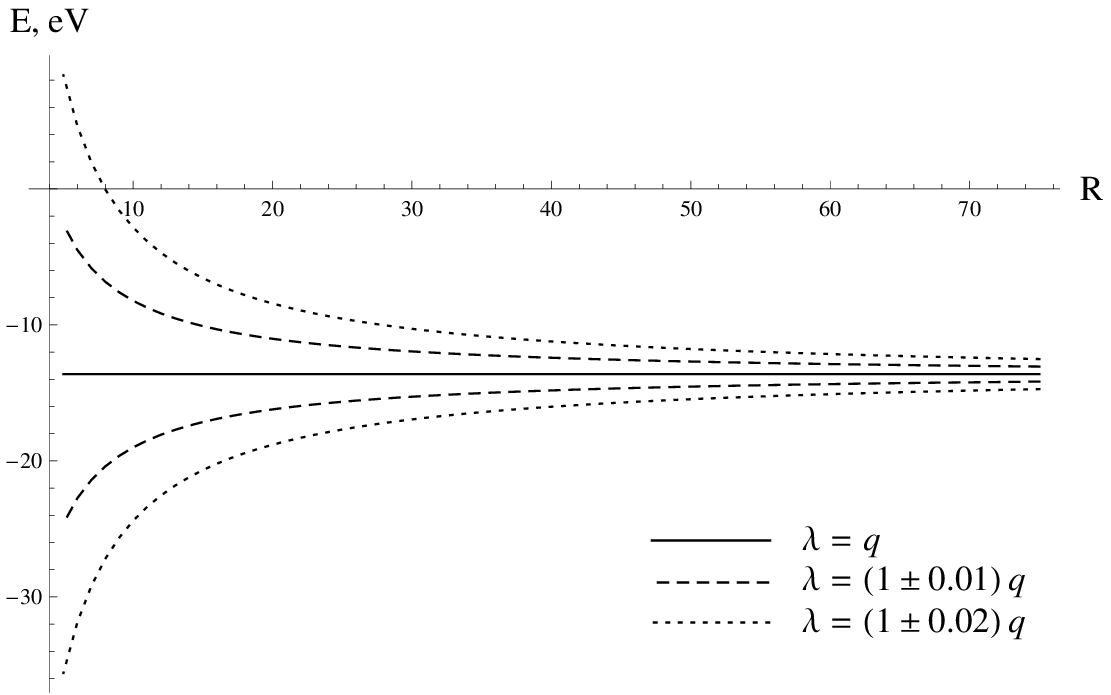}\\
\caption {The lowest $s$-level of  atomic H with $q=\a $
in a cavity with boundary conditions of ``not going out''
(\ref{f3})  as a function of radius $R$. The sign of the shift of $\l$ relative to $q$ and
corresponding shift of levels relative to $E_{1s}$ coincide. }
\label{fig:Hydrogen-Shr}
\end{figure}

The analogy between a particle in a well and   H  in a
cavity remains valid  for $R \to \inf$ too, where it could be
easily checked  by means of asymptotic expansion for $\F_R \ , \
\F_R(b+)$ in (\ref{f19}), that in the case of surface attraction
$\l <0$ there exists one more level $\tE (R)$  with negative  limiting value $\tE(\inf) = -
\l^2/2$, besides the discrete
spectrum of the free atom,  and  power asymptotical behavior for   $R \to \inf$
\beq
\label{f27} \tE (R) \to - \l^2/2 -  (q-\l)/R + O(1/R^2) \ , \ R \to
\inf \ . \eeq
For  $\l < -q < 0 $ this analogy could be extended on the
whole range of cavity sizes, since under these conditions $\tE
(R)$ turns out to be the lowest atomic $s$-level  with the form of
shifted downwards hyperbole, as for a particle in a well.

Now let us turn to the next type of atomic levels in a cavity,
which appear under assumption, that   $\g R$ remains finite for $R
\to 0$. To maintain the connection with two previous types of levels,
we consider only s-levels with $l=0$ and rewrite (\ref{f19}) in the form
\beq \label{f29} \left. \( 2 \pd /\pd z + \l/\g -1\)
\F(b,2,z)\right|_{z=2\g R} =0   \ . \eeq
Since $\g \to \hbox{Const} /R \ $ for $R \to 0$, then $\l/\g \to 0 \ , \
b_0=1-q/\g \to 1 \ ,$ and so (\ref{f29}) transforms into \beq
\label{f30} \left. \( 2 \pd /\pd z -1\) \F(1,2,z)\right|_{z=2\g R}
=0  \ . \eeq Taking account of $\F(1,2,z)=\(e^z -1\)/z \ $, from
(\ref{f30}) one obtains \beq \label{f31} \g R=i x_n \ , \quad \tan
x_n=x_n \ , \eeq what describes  positive energy levels with
the asymptotics
 \beq \label{f32}
E_n(R) \to {x_n^2 \over 2 R^2} \ , \quad R \to 0  \ , \eeq i.e.
excited  states of a particle (electron) in a well with Neumann
boundary conditions (\ref{f6}). So    all the $s$-levels besides
$1s$ (provided that the latter turns out to be the lowest one and falls down
for $R \to 0$, what implies  $|\l|< q$) should  for $R \to 0$
reveal asymptotical behavior  (\ref{f32}), while levels with $l
\not=0$  lye even higher due to the centrifugal term. At the
same time,   for $R \gg 1$ all the $ns$-levels  (as well as levels
with $l \not=0$) tend to their asymptotical values, corresponding
to those of the free atom, exponentially fast
 \beq \label{f33}
E_n(R) - E_n \to  \[ {\g_n \over n! } \]^2 \ { \l -\g_n \over \l
+\g_n} \ \( 2 \g_n R\)^{2n} \ e^{-2\g_n R} \ ,
\nonumber \\ \eeq
\beq \label{f33} \g_n R \gg 1 \ , \eeq
where \beq \label{f34}
E_n= - \g_n^2 / 2 \ , \quad \g_n = q / n  \ , \quad n=1,2,\dots \ , \eeq
are the $ns$-levels of the free atom. Remark, that levels with $\g_n < \l$ should approach their asymptotics from above, while those with $\g_n > \l$ from below.

It should be specially noted, that the asymptotics (\ref{f33})
turns out to be an exceptional feature of those confined atom
levels, which originate from the discrete spectrum of the free
atom, since such asymptotics is created  by approaching the
argument of the factor $\G^{-1}(b)$, entering the asymptotics  of
the Kummer function $\F (b,c,z)$, to the pole $b \to -n_r, \
n_r=0,1,\dots.$ Asymptotics for $R \to \inf$ of all the other
atomic levels in a cavity, which originate from the continuous
spectrum of the free atom and the additional level (\ref{f27}), caused
by attractive interaction with environment, turns out to be  a
power series in $1/R$, and their  asymptotical values could be
either non-negative only, or for $\l <0$ contain one negative
point $\tE (\inf)=-\l^2/2$.

If  $\l = \pm \g_n$, the asymptotics (\ref{f33})  modifies in the
next way. The exponential behavior is preserved, while
the non-exponential factor undergoes changes in such a way, that
the $ns$-levels approach their asymptotics of the free atom from above
only. For  $\l =\g_n >0$ their asymptotics takes the form
 \beq \label{f35}
E_n(R) - E_n \to  (n-1) \[ {\g_n \over n! } \]^2  \ \( 2 \g_n
R\)^{2(n-1)} \ e^{-2\g_n R} \ ,
\nonumber \\ \eeq
\beq \label{f35} \g_n R \gg 1 \ , \eeq
while for the lowest level  $E_1(R)$ the exponential part disappears completely, since in this case $\l=\g_1=q$, and as it was mentioned above,  $E_1(R)$ becomes a constant, which coincides with  $E_{1s}=-q^2/2$.

For $\l =-\g_n < 0$ instead of (\ref{f33}) one obtains
 \beq \label{f36}
E_n(R) - E_n \to { 1 \over n+1} \  \[ {\g_n \over n! } \]^2  \ \(
2 \g_n R\)^{2(n+1)} \ e^{-2\g_n R} \ ,
\nonumber \\ \eeq
\beq \label{f36} \g_n R \gg 1 \ , \eeq and
moreover, the limiting point $\tE (\inf)$ of the level $\tE (R)$ with the power
asymptotics (\ref{f27})  coincides with the corresponding level $E_n$ of the
free atom (\ref{f34}), what in turn represents a remarkable example
of von Neumann-Wigner avoiding crossing effect, i.e. near levels reflection  under perturbation \cite{LL},\cite{NW}
--- infinitely close to each other for $R \to \inf$ levels
$E_n(R)$ and  $\tE (R)$  should for decreasing $R$ diverge in
opposite directions from their common limiting point  $E_n$.
Perturbation in this case is performed by the atomic nuclei
Coulomb field, since  under general boundary conditions (\ref{f3}) the electronic wavefunction doesn't vanish on the
cavity boundary, and so  for $R \gg 1$ the maximum of electronic
density should be shifted into the region of large distances
between the electron and nuclei, where the contribution of the
Coulomb field is negligible compared to boundary effects. When
$R$ decreases, the Coulomb field increases, hence   $E_n(R)$
should go upwards according to (\ref{f36}), while $\tE (R)$ goes
downwards according to the asymptotics \beq \label{f37} \tE (R) \to
E_n -  {n+1 \over n} \ {q \over R} + O(1/R^2) \ , \ R \to \inf \ .
\eeq

 So the energy spectrum of atomic H  (with $q>0$), confined in a cavity with Robin's condition (\ref{f3}), turns out to be the following. For $\l=q$ the lowest  $s$-level acquires the constant value $E_{1s}$ of the free atom, for $\l>-q$ it behaves for  $R \to 0$ according to (\ref{f26}) with an energy shift depending on $\sign \(\l -q\)$ and for  $R \gg 1$ it approaches $E_{1s}$ exponentially fast, while for $\l \leq -q <0$ it transforms into the level  $\tE(R)$ with power asymptotics (\ref{f27}). Excited states in all the cases should for  $R \to 0$ reveal the behavior (\ref{f32}). And for an H-like atom there once more takes place the situation, similar to that for a particle in a potential well, namely  --- whenever $\l <q$, the  atomic state with largest bound energy, which could sufficiently exceed the bound energy of the lowest level of the free atom  (\ref{f23}),  takes place in the smallest cavity.

\subsection*{4.  Atomic H in the Wigner-Seitz cell}

So far, by formulating the  confinement problem (\ref{f2}-\ref{f3})
it was implied, that a particle in  such a ``not going out'' state
interacts with environment only on the cavity boundary $\S$, i.e.
through certain $\d$-like potential, what leads to the surface term in the
energy functional (\ref{f1}). In a more realistic approach one
should consider instead of a $\d$-like  interaction an outer potential shell
of nonvanishing  thickness $d$, into which the particle
penetrates and interacts there with cavity environment. In the
limit $d \to 0$ such potential shell should transform into contact
interaction on the surface $\S$. For these purposes the boundary
condition (\ref{f3}) should be replaced  by an equation of
Schroedinger type, describing particle interaction with medium
inside the shell, whose potential might be quite different from
$U(\vec r)$. In the case of
spherical cavity and shell the first choice for the shell potential
is a constant $U_0$, as by modelling the endohedral environment \cite{Connerade1}. Then instead of
(\ref{f3}) one obtains \beq \label{f38}
 \left[ -{\h^2 \over 2m } \D + U_0  \right] \p=E \p \ , \quad R \leq r < X=R+d \ ,
\eeq with Neumann condition on the outward
shell boundary at $X=R+d$ \beq \label{f39} \left.  \pd   \p /\pd r \right|_{r=X} =0 \ . \eeq There is no $\l$ in (\ref{f39}), since the role of interaction with environment is played
now by eq. (\ref{f38}). Moreover, as it was mentioned above, such approach allows for a
sufficiently more wide problem statement, since it doesn't imply
genuine  trapping of a particle by the given volume. In
particular, boundary condition  (\ref{f39}) appears in a quite
natural way by considering the particle ground state in a cubic lattice, formed
by cavities of the same type in a crystal matrix, within the
Wigner-Seitz model \cite{WS} with the same assumption, that the genuine wavefunction of the problem will actually have the highest $(0^h)$ crystallographic symmetry which is not very far from the spherical one. The well-known example of such
sublattices is given by interstitial sites in  certain metals and
alloys \cite{Alefeld}-\cite{PdHx}. The cavity together with an  outer shell form in
this case a kind of the Wigner-Seitz cell, while (\ref{f39}) turns out to be  the condition of periodic continuation of the particle
wavefunction between neighboring cells.

It is also worth while
noticing, that in the latter case instead of the  energy level with such
periodic wavefunction the whole set of states $\p_{\vec k}(\vec
r)=u_{\vec k} (\vec r) \exp\(i{\vec k} {\vec r}\)$ with $u_{\vec
k} (\vec r)=u_{\vec k} ({\vec r} +{\vec a})$, where $\vec a$ is
the period of the cavities sublattice, while the wavevectors $\vec k$
fill in the corresponding  first  Brilluen zone,   should be
considered. In any case, however, periodic wavefunction  describes
the level with ${\vec k}=0$, hence the position of the bottom of
the first Brilluen zone, what is quite important itself.

Besides this, the magnitude of the shell potential $U_0$ should
depend on the penetration depth $d$ in such a way, that provides a
transition to a $\d$-like interaction with coupling constant  $\l$
for $d \to 0$, what for \beq \label{f40} U_0 d \to {\h^2 \over 2m}
\ \l  \ , \quad d \to 0 \ . \eeq Note, that the limits $d \to 0$
and $U_0 \to \infty$ don't commute --- when $U_0 \to \infty$ with $d
\not= 0$, then one obtains  confinement in a cavity by an
impenetrable barrier, hence with boundary condition (\ref{f7}),
while for $d \to 0$ and finite product $U_0 d$, on the contrary,
the case of general  boundary condition (\ref{f3})
takes place.

For $s$-levels of  atomic H in such a cavity with an outer
shell instead of  contact interaction, one obtains the following
spectral problem (in units, introduced in sect. 3)
\beq \label{f41}
\[ \k R (1-\k X \th \k d ) +(q/\g -\g R) (\th \k d - \k X)\]\F_R +
\nonumber \\ \eeq
\beq \label{f41}
 + (1-q/\g) (\th \k d - \k X) \F_R(b+)=0 \ , \eeq
where $\k^2=2(U_0-E) \ $, while all the other quantities are
defined  as in (\ref{f18}) and (\ref{f20}). It is easy to see, that
the relation (\ref{f40}) gives \beq \label{f42} \k R {1-\k X \th \k
d \over \th \k d - \k X} \to \l R-1 \ , \quad d \to 0 \ , \eeq
whence it follows, that for $d \to 0$ eq.(\ref{f41}) transforms
into eq.(\ref{f19}) for atomic $s$-levels with boundary condition
(\ref{f3}).

It should be specially remarked, that the limits $d \to 0$ and $R
\to 0$ don't commute either. In particular, if  $d \not= 0$, then
for the lowest $s$-level the solution of eq.(\ref{f41}) for $R \to
0$  leads to \beq \label{f43} E_{ground} (R) \to U_0 \ , \quad R
\to 0. \eeq The latter could be easily detected from (\ref{f11}), which in this case gives
\beq \label{f44} E_{ground}
(R) < E_{trial}(R)=
\nonumber \\ \eeq
\beq \label{f44}
 ={3 R^2 + 3 R d + d^2 \over (R+d)^3} \ U_0 d -
{3 R^2 \over 2 (R+d)^3} \ q \ , \eeq
and $E_{trial}(R \to 0) \to
U_0 \ ,$ combined  with the above-mentioned  feature, that for $R
\to 0$ the estimate (\ref{f11}) turns out to be exact.

More precisely, there are two types of solutions of eq.(\ref{f41})
for $R \to 0$.  The first one originates from (\ref{f41}) by
neglecting the term with $\k R$ and omitting the common factor
$(\th \k d - \k X) \to (\th \k d - \k d)$, what leads to the
following relation \beq \label{f45} (q/\g -\g R)\F_R + (1-q/\g)
\F_R(b+)=0 \ . \eeq For $R \to 0$ eq.(\ref{f45}) contains no
solutions with finite energy, since when $\g R \to 0$, then $\F_R
\ , \ \F_R(b+) \to 1$, hence (\ref{f45}) reduces to $1=0$, and
otherwise, when  $\g R \to \hbox{Const} \not=0$, then $q/\g \to 0$
and $b_0 \to 1$, thence (\ref{f45}) could be simplified up to \beq
\label{f46} z \F(1,2,z) =2 \F(2,2,z) \ . \eeq
  Eq.(\ref{f46}) in turn reduces to $e^z +1=0$, whence $\g_n=i (\pi/2 + \pi n)/R \ $, what corresponds to a series of highly excited $s$-states with energies
 \beq \label{f47}
E_n \to {(\pi/2 + \pi n)^2 \over 2 R^2 } \ , \quad R \to 0 \ . \eeq
The second type of solutions of (\ref{f41}) for $R \to 0$ emerges
from the factor $(\th \k d - \k d) \ $, what gives $\k_n=i x_n/d$
with $x_n$ being the solutions of eq. $\tan x_n=x_n$, and so leads
to another series of $s$-levels, corresponding to the energy
spectrum of a particle in a well of radius $d$ and Neumann
boundary condition (\ref{f6},\ref{f39})
 \beq \label{f48}
E_n=U_0 +{ x_n^2 \over 2 d^2 }  \ . \eeq These levels reveal
a finite limit for  $R \to 0$, while the lowest one, corresponding
to $x_0=0$, meets  the limiting value  $E_0(R \to 0)=U_0$.

 It is easy to verify,
that there are no solutions of eq.(\ref{f41}) for $R \to 0$ besides
(\ref{f47}) and (\ref{f48}). So   the effect of infinite descent  of
the lowest level for $R \to 0$,  which  takes place  in the case of
contact surface interaction for $\l < q$, doesn't occur for the
potential shell of nonvanishing width.

The physical meaning of series (\ref{f47},\ref{f48}) should be quite
clear. The levels (\ref{f47}) correspond to $s$-states of
continuous spectrum of the free atom, when the latter is confined
in a cavity with $R \to 0$, while the levels (\ref{f48}) originate
from  $ns$-levels with exponential asymptotics (\ref{f33})  and a
finite number of levels $\tE_k$ with power asymptotics for $R \to
\inf$, which appear for  $U_0 <0$ and  turn out to be direct analogies
of $\tE(R)$ for the case of contact
interaction with  $\l< 0$ (\ref{f27}) .

Compared to the case of contact interaction (\ref{f33}), the
asymptotics of $ns$-levels for   $R \gg 1$ is modified in the
following way
 \beq \label{f50}
E_n(R) - E_n \to
\nonumber \\ \eeq
 \beq \label{f50}
\to \[ {\g_n \over n! } \]^2 \ { \k_n \th \(\k_n d \)
-\g_n \over \k_n \th \(\k_n d \) +\g_n} \ \( 2 \g_n R\)^{2n} \
e^{-2\g_n R} \ ,
\nonumber \\ \eeq
 \beq \label{f50} \g_n R \gg 1 \ , \eeq
 where
 \beq \label{f51}
\k_n = \sqrt {2 U_0 + \g_n^2} \ , \eeq while $E_n$ and $\g_n$ are
defined as in (\ref{f34}). It follows from  (\ref{f50}), that for
 \beq \label{f52}
|\k_n \th \(\k_n d \)| < \g_n \eeq the curves $E_n(R)$  approach
the  $ns$-levels of the free atom (\ref{f34}) for $R \gg 1$ from
below, while for $|\k_n \th \(\k_n d \)| > \g_n$ from above.
Therefore the curves $E_n(R)$ could for finite $R$ reveal
nontrivial minima, which lye below  the corresponding  $ns$-levels
of the free atom (\ref{f34}), provided that  the  relation (\ref{f52}) is
satisfied. The specific feature of the problem with an outer shell
is that now such a minimum, and the deepest one, exists for  the
lowest $s$-level as well, whereas for vanishing width of the shell
and $\l < q$ this level should for $R \to 0$  reveal an infinite
falldown, and so nontrivial minima could appear for excited
states only. A crude estimate for such a minimum for the lowest
$s$-state can be received from inequality (\ref{f43}) by solving
$\partial E_{trial}(R) / \partial R |_{R_0}=0$, what gives
\beq \label{f53} R_0={2 q d \over q - 2 U_0 d } \ , \eeq and \beq
\label{f54} E_{trial}(R_0)= { 1 \over d } \  { 2 U_0 d - q \over
\(2 U_0 d - 3 q \)^3 } \times
\nonumber \\ \eeq
\beq \label{f54}
\times  \[ 4 \(U_0 d\)^3 - 16 q \( U_0 d \)^2 +
19 q^2 U_0 d - 6 q^3 \] \ . \eeq
The eq. (\ref{f53}) predicts the
existence of a nontrivial minimum for the lowest state only when
$2 U_0 d < q$, what is more crude, than the exact relation
(\ref{f52}). The difference, however, should be quite clear, since
the estimate (\ref{f11}) works well only for small cavities of such
type, hence for small $d$, when $2 U_0 d$ should be identified
with $\l$ and so $2 U_0 d < q$ is nothing else, but  the relation
$\l < q$.

As for the boundary condition (\ref{f3}), the asymptotics
$\tE_k(\inf)=\tE_k$ of power levels $\tE_k(R)$ with negative
limiting values for $R \to \inf$ is found from (\ref{f41}) by
taking account of the main exponential term in the asymptotics of
the Kummer function, what yields the following relation \beq
\label{f55} \tk_k \th \(\tk_k d \) + \tg_k=0 \ , \eeq where
$\tk_k=\sqrt {2 U_0 + \tg^2_k} \ , \quad \tE_k=-\tg^2_k/2.$ Note, that if $\tg_k=\g_n$, i.e.
the levels $\tE_k$ and $E_n$ possess the same limiting value for $R \to \inf$, then the l.h.s. of (\ref{f55})
coincides with the denominator in the asymptotics of exponential levels (\ref{f50}). So
 vanishing denominator in (\ref{f50}) implies once more the change in the asymptotical behavior of the exponential level due to the Neumann-Wigner reflection effect, what is discussed in detail for the case of the lowest level  below.

It follows from  (\ref{f55}), that such power levels with $\tE_k <
0$ might appear only for $U_0<0$, when (\ref{f55}) takes the form
\beq \label{f56} \sqrt{2 |U_0| -\tg^2} \tan \( \sqrt{2 |U_0|
-\tg^2} \ d \) = \tg \ , \eeq
and  is nothing else but the
equation for even levels in one-dimensional square well of width
$2d$ and depth $U_0$. Therefore the levels $\tE_k$ exist for any
$U_0<0$ and $d>0$, their values lye in the interval $U_0<\tE_k<0$,
while their total number $K$ is defined from $\pi (K-1) < 2|U_0| d \leq
\pi K \ , \ K=1,2, \dots $.

The asymptotics of the levels $\tE_k(R)$ for $R \to \inf$ takes the
form \beq \label{f57} \tE_k (R) \to   \tE_k -  { q \over R} \ {
|U_0|+\tE_k \over |U_0| \(1+ \tg_k d \) } - {1 \over R} \ { \tg_k
\over 1+ \tg_k d}  + O(1/R^2) \ ,
\nonumber \\ \eeq
\beq \label{f57}  R \to \inf  \ . \eeq
Note, that in such a problem there exist other power levels with $\tE_k
> 0$ , which correspond to imaginary wavenumbers $\tg_k$ and so
should be found from the asymptotics of the Kummer function
including the power term besides the exponential one, but these levels
lye wittingly higher, than the power (\ref{f57}) and exponential
(\ref{f50}) ones, whereas  our main interest is first of all bent
on the lowest atomic levels.

Another crucial difference between power $\tE_k (R)$ and
exponential $E_n(R)$ levels is that  the origin of the formers is the  attractive interaction between the
particle (atomic electron) with cavity boundary (outer shell), rather than   the interaction with the inner shell (atomic nuclei).
In fact, (\ref{f55}) is equivalent to the equation, defining
the $s$-levels of a particle in the attractive potential of a spherical shell
\beq \label{f58} U(r)=U_0 \ \tt \( R \leq r \leq R+d \) \eeq with
Neumann boundary condition (\ref{f39}) on the
outward boundary, in the limit $R \to \inf$. In a slightly different language, this circumstance has been pointed out in \cite {Connerade1, Wiese}.

A more detailed analysis of eq.(\ref{f41}) turns out to be most
conveniently performed by means  of its numerical solution for a
concrete set of parameters $U_0$ and $d$,  corresponding to
realistic scales of microcavities, in which such a ``confined'' H state could occur (to simplify the discussion, henceforth
we'll deal with energy in eV). For $|U_0|$ this is $1 - 100$ eV,
for $d$ --- fractions of the Bohr radius $a_B=1/\a \simeq 137$,
more concretely $d=x a_B$ with $x=2^{-p}, \ p=-1,0,1,...,4$, where the largest  $2 a_B$ is chosen according to the mean
width of one-atom surface shell, while the smallest  $a_B/16$  ---
in accordance with the lower limit, following from relativistic effects \cite{PEPAN}.
The range of values  for $|U_0|$ is defined by taking into account,
that $|U_0|$ could vary from $\sim 1$ eV for vacuum ``bubbles'' in
superfluid  He$^4$ \cite{Maris} up to dozens eV
in  quantum chemistry \cite{Jask}-\cite{Sako}.

The most simple and transparent  example is given by the potential
barrier $U_0 >0$. In this case the lowest energy atomic H state
coincides with the exponential $1s$-level, whose behavior as a
function of $R$ for $U_0=10$ eV  is shown  on Fig.2 (for $R\geq 10$).
\begin{figure}
\includegraphics[width=8.6 cm]{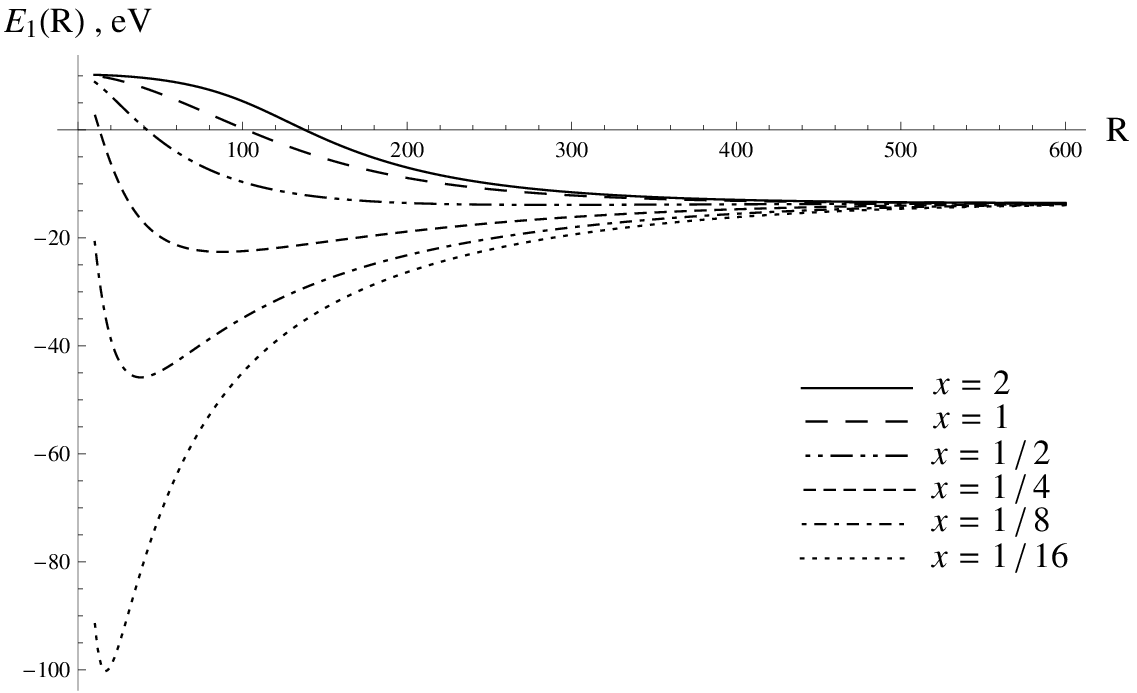}\\
\caption {The lowest $1s$-level of atomic H with  $q=\a$
as a function of cavity radius $R$ for $U_0=10 \ $eV and $d= x
a_B$.}
\label{fig:U0=10 eV}
\end{figure}
In accordance with relation
(\ref{f52}), which in this case gives  $d < 100$, there are
pronounced minima for such $U_0$ and $d=a_B/4 \ , \ a_B/8 \ , \
a_B/16$  on the curves $E_1(R)$, and so a cell with such parameters turns out to be an effective H-trap. A minimum exists also for $d=a_B/2$, but here it is
quite weak (bound energy exceeds only $\simeq 13,9$ eV) and
takes place at $R=289$, while for $d=a_B \ , \ 2a_B$, on the
contrary, the H state with the lowest energy is achieved for
infinitely increasing $R$.

The behavior of the lowest level reveals a more pronounced
dependence on $d$, as well as on $U_0$,  in the case of attraction
in the outer shell ($U_0 < 0$) due to increased  amplitude of
electronic wavefunction  near the outward boundary. In particular, for
$U_0=-10$ eV  (Fig.3)  the relation (\ref{f52}) is
fulfilled for all $d$, hence  nontrivial minima in the bound
energy exist now for all $d$, including   $d=2 a_B$, when a
minimum with bound energy $14,8$ eV is achieved at $R=254$.
\begin{figure}
\includegraphics[width=8.6 cm]{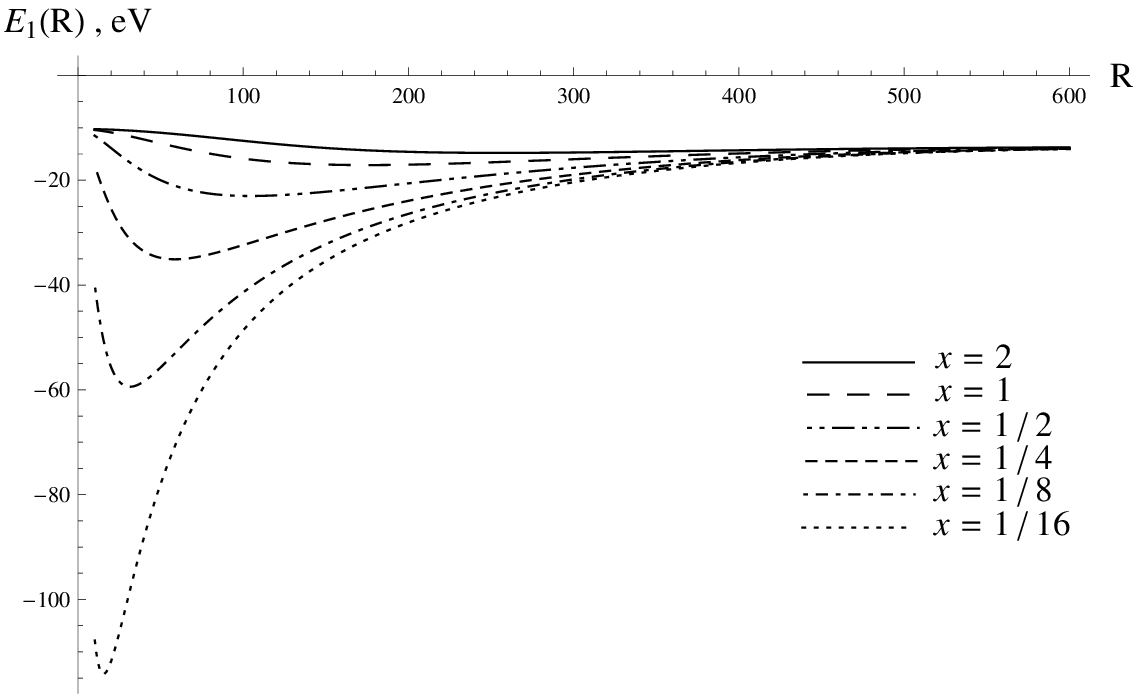}\\
\caption {The lowest $1s$-level of atomic H with  $q=\a$
as a function of cavity radius $R$ for  $U_0=-10 \ $eV and $d= x
a_B$.}
\label{fig:U0=-10 eV}
\end{figure}

Now let us consider the most interesting case of energy levels
reconstruction for  atomic H  in a cavity with such an
attraction in the  outer shell, that the lowest atomic level
turns out to be the power one $\tE_1(R)$. For these purposes the
``critical'' potential  $U_{0 \ \hbox{\footnotesize crit}}$, which provides the
coincidence of $\tE_1 (\inf)$ with the limiting value  of the
first exponential level $E_1(\inf)=E_{1s}$, what implyes \beq
\label{f59} \tg_1=\g_1 \  , \eeq should be firstly determined.

The  form  of  $U_{0 \ \hbox{\footnotesize crit}}$ as a function of the shell width
$d$ is shown on Fig.4. Above $U_{0 \ \hbox{\footnotesize crit}}(d)$ there lyes the
region of $U_0$ and $d$, where the lowest level turns out to be
the exponential $E_1(R)$, while below --- the region, where the power $\tE_1 (R)$ is the
lowest.

\begin{figure}
\includegraphics[width=8.6 cm]{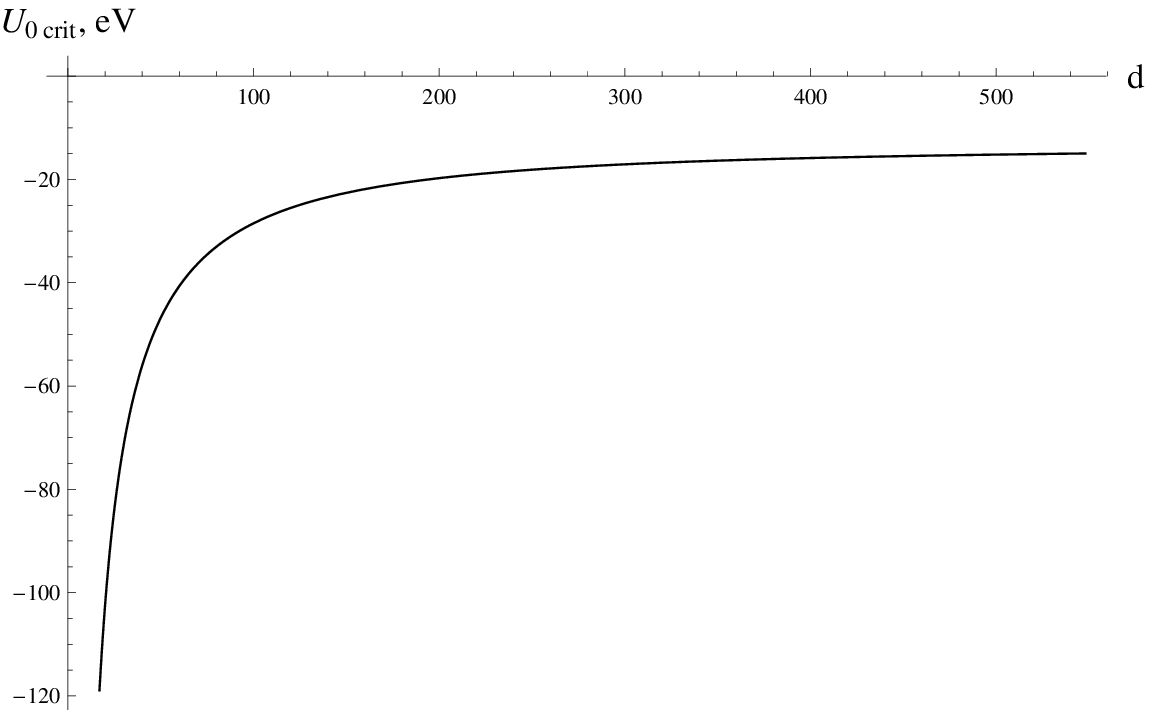}\\
\caption {The  ``critical'' potential  $U_{0 \ \hbox{\footnotesize crit}}$   as a
function of $d$.}
\label{fig:U0crit}
\end{figure}

It is easy to find  from eqs.(\ref{f55}, \ref{f59}), that  for $d \to
\inf$ the limiting value of $U_{0 \ \hbox{\footnotesize crit}}(d)$ should coincide with
the lowest level of the free H $(\ref{f23})$ with the following
asymptotics \beq \label{f60} U_{0 \ \hbox{\footnotesize crit}}(d) \to E_{1s} - {\pi^2
\over 8 d^2} + \ { \pi^2 \over 4 q  d^3} + O(1/d^4) \ , \quad d
\to \inf \ , \eeq while for $d \to 0$ the  ``critical'' potential
decreases according to \beq \label{f61} U_{0 \ \hbox{\footnotesize crit}}(d) \to - q /
2d \ , \quad d \to 0 \ . \eeq The numerical values of $U_{0 \ \hbox{\footnotesize crit}}$ for $d$ under consideration are presented in Tab.1 (besides
$d=a_B/16$, since in this case $|U_{0 \ \hbox{\footnotesize crit}}|$ turns out to be
too large).

\beq
{\footnotesize
 \begin{array}{c|c|c|c|c|c|c|c}
 x=d/a_B & 1/8 & 1/4 & 1/2 & 1 & 2 & 4 & 16 \\
 \hline
 U_{0 \ \hbox{\footnotesize crit}}, eV & -118.9 & -64.2 & -37.0 & -23.7 & -17.6  & -15.0 &
   -13.7  \\
\end{array}}
 \notag \eeq
{\small TAB. 1: The values of $U_{0 \ \hbox{\footnotesize crit}}$ for $d=x a_B$.}
\vskip 0.3 cm

The behavior of the curves $E_1(R)$ and $\tE_1(R)$ for $U_0=0.9 \ U_{0 \ \hbox{\footnotesize crit}}$ and  $U_0=U_{0 \ \hbox{\footnotesize crit}}$ is shown on Fig.5 for $d=a_B$. It is easy to recognize here the effect of avoided crossing, discussed in \cite{Connerade1,Wiese}, which shows up now in the change of asymptotics of the lowest level, when the value of $U_0$ coincides with $U_{0 \ \hbox{\footnotesize crit}}$.
\begin{figure}
\includegraphics[width=8.6 cm]{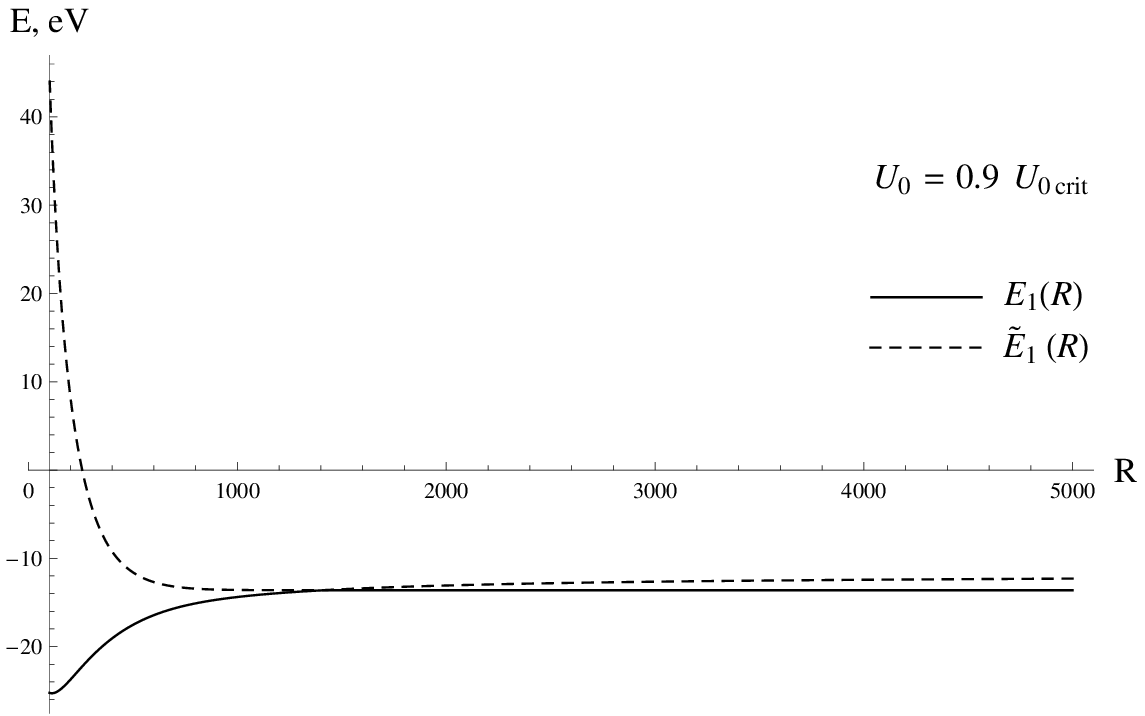}\\
\includegraphics[width=8.6 cm]{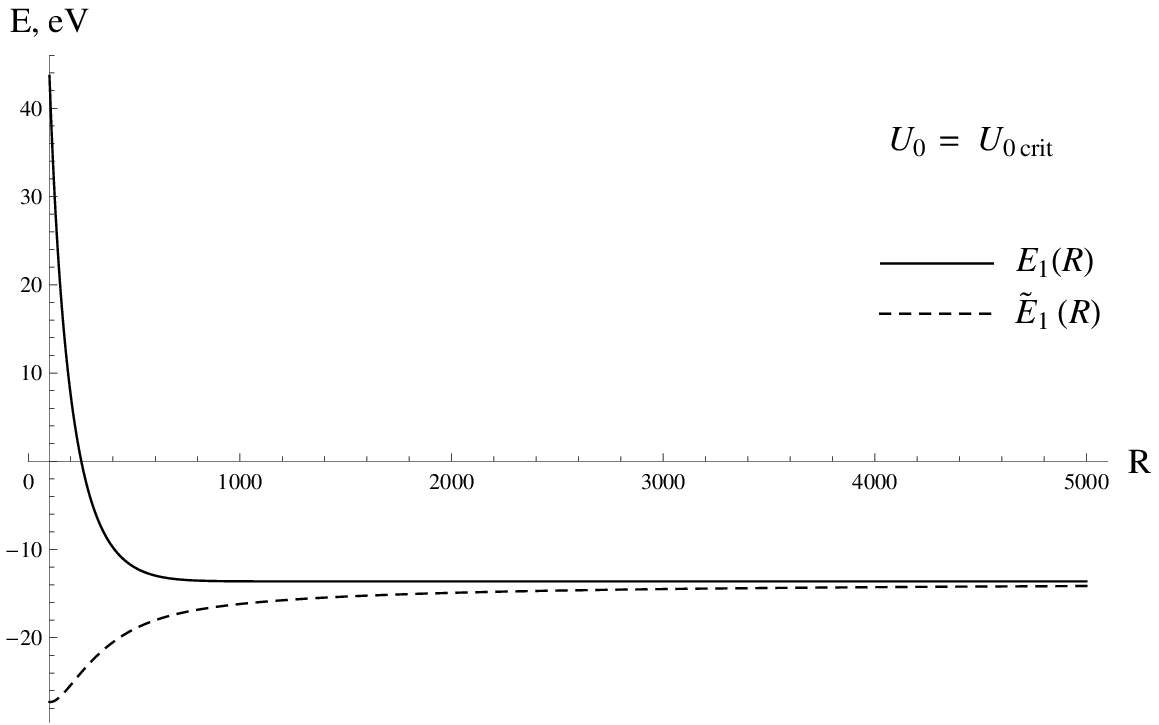}\\
\caption {$E_1(R) \ , \tE_1(R)$ for   $U_0=0.9 \ U_{0 \ \hbox{\footnotesize crit}}$ and $U_0=U_{0 \ \hbox{\footnotesize crit}}$ for $d=a_B$.}
\label{fig:E1,E1tilde}
\end{figure}

It should be emphasized, that the change of the lowest level
asymptotics for $R \to \inf$ from exponential into the power one
takes place indeed when $U_0$ reaches $U_{0 \ \hbox{\footnotesize crit}}$ from above,  not
earlier and not later. In this case the limiting point $E_{1s}$ of the free
atom is the same for the curves $E_1 (R)$  and $\tE_1 (R)$, and so
as  for the case of contact interaction
(\ref{f36},\ref{f37}),  the exponential level $E_1 (R)$ should approach
its limiting point from above, thence the power $\tE_1 (R)$
turns out to be the lowest one due to the Neumann-Wigner reflection.
Let us underline specially, that it is indeed an exchange of
asymptotical behaviour for the lowest level --- the levels $E_1
(R)$  and $\tE_1 (R)$ could be infinitely close to each other, but
don't touch and all the more that intersect, since all the
$s$-levels in such a problem  cannot be
degenerate. Note also, that  this exchange of asymptotical behavior proceeds in the following way --- when $U_0 \to U_{0 \ \hbox{\footnotesize crit}}$ from above, the denominator in the r.h.s. of $(\ref{f50})$ tends to zero, hence the exponential tail of the curve $E_1(R)$ is shifted to more and more large $R$, for $U_0=U_{0 \ \hbox{\footnotesize crit}}$ it disappears completely, and so the power behavior extends to the whole half-axis $0< R < \inf$.

The behavior of the level $\tE_1 (R)$ for $U_0=U_{0 \ \hbox{\footnotesize crit}}(d)$,
i.e. at the  moment when it becomes the lowest
one, is shown  on Fig.6 for  $d=2^{-p} a_B$ with $p=-1,0,...,3$.
(The case $d=a_B/16$ is omitted, since
$U_0$ and the upper limit of bound energy acquire in this case too
large values $ U_0 \sim - 240$ eV,  bound energy $ >300$ eV, and so
in background of the curve for $d=a_B/16$  the details of the
other curves become illegible.)

\begin{figure}
\includegraphics[width=8.6 cm]{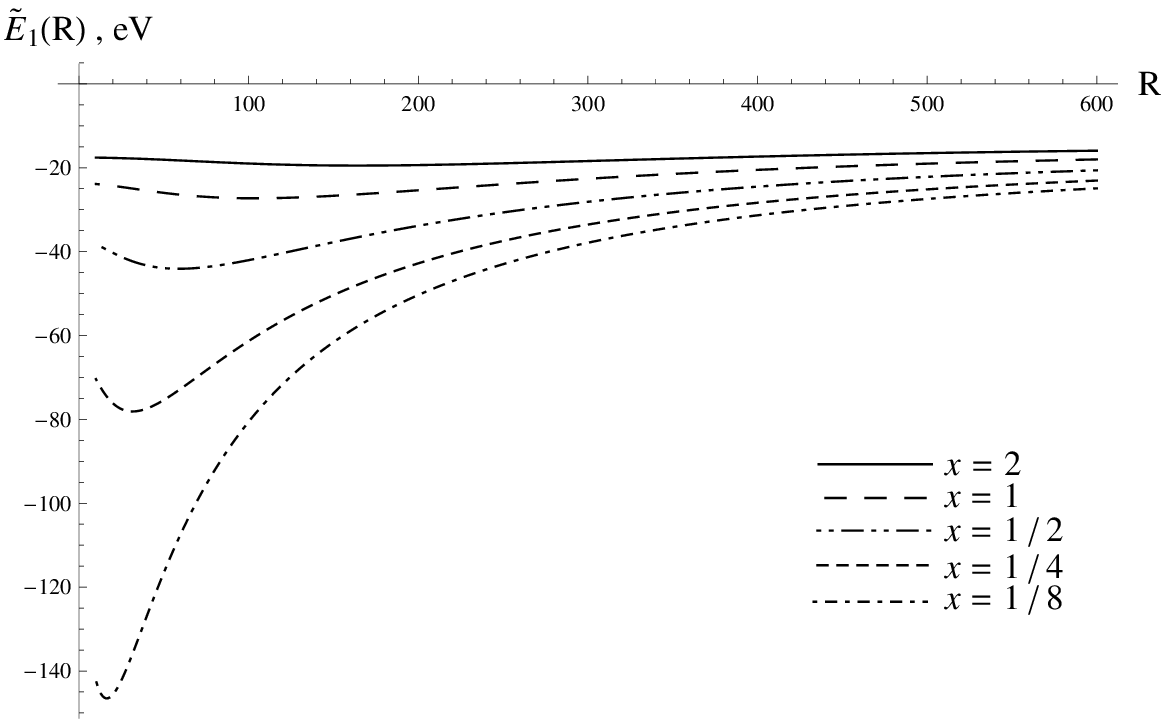}\\
\caption {The behaviour of  $\tE_1(R)$ for $U_0=U_{0 \ \hbox{\footnotesize crit}}(d)$ and  $d=x a_B$.}
\label{fig:Etilde}
\end{figure}

Fig.6 shows explicitly the power asymptotics of the level $\tE_1
(R)$ for  $R \to \inf$. The shift of power levels relative to
their asymptotical value decreases sufficiently slowlier than
that of the exponential ones, which arrive at their asymptotics
for $R$  of order of several $a_B$ already, and so such a ``confined'' atomic H state
turns out to be energetically
favorable up to actual nanoscales, provided that $U_0 \leq U_{0 \ \hbox{\footnotesize crit}}(d)$. Numerical values (in eV) for
the shift of bound energy of the power level $\tE_1 (R)$ relative
to the free H are given in Tab.2 for a cavity of nanosize with
$U_0= U_{0 \ \hbox{\footnotesize crit}}(d)$, i.e. at the moment when $\tE_1 (R)$
transmutes into the atomic H ground state, for $d=a_B, \ a_B/2, \
a_B/4$.

\beq
{\small
\begin{array}{c|c|c|c|c}
 R & 1  \ nm & 10 \ nm & 100 \ nm & 1000 \ nm    \\
 \hline
 \D E(d=a_B)  & 1.00894 & 0.102517 & 0.0102667 & 0.00102686   \\
 \D E(d=a_B/2) & 1.5722  & 0.157199 & 0.0157202 & 0.00157203     \\
\D E(d=a_B/4) & 2.08505 & 0.20661   & 0.0206425 & 0.00206407
   \end{array} }
   \notag \eeq
{\small  TAB. 2: The values of $\D E=E_{1s}-\tE_1 (R)$   in a nanocavity with  $U_0=U_{0 \ \hbox{\footnotesize crit}}(d)$,
when $\tE_1 (R)$ transmutes into the lowest atomic level. }

\subsection*{5. Conclusion }
To conclude let us firstly mention, that a single cavity might be just a simple hollow cage without any special confining property, but  a large set of cavities, forming a cubic lattice, could reveal such properties due to  quantum coherence  effects, similar to those creating the metallic bond in the Wigner-Seitz model \cite{WS}. As a result, each single cavity of such a lattice transforms into a kind of the Wigner-Seitz cell, formed by a cavity with an outer potential shell.

The properties of the particle state in such a cell
turn out to be quite different from confinement
by a potential barrier \cite{Aquino}, \cite{Sen1}. An example of such kind is presented in sect.4 by atomic H in a cavity with potential shell, provided that the set of cavities, occupied by atoms, forms a crystal structure similar to that of an alcaline metal. In particular, in dependence on the outer
shell parameters the upper limit for the bound energy of H  in such a cell could be more large, than several times the
bound energy of the lowest $1s$-level of the free atom. At the
same time, in the case of a power lowest level the bound energy decreases very slowly for increasing cavity size, therefore such a  state should be energetically
favorable compared to the free atom  up to actual nanocavities with $R \sim 100-1000$ nm. The latter circumstance  means, that artificial macroscopic
lattices, created from such nanocavities in suitable media, could serve as quite effective containers of H. Even more
interesting results for searching possible new effects appear in
the case of more complicated atoms and simplest diatomic molecules in
such a cell and a lattice  with the same
parameters as employed in sect.4 \cite{Sv-Tol}.

\vskip 1 true cm

\begin{acknowledgments}
 One of us (K.S.) is grateful to  Dr. Maxim Ulybyshev from ITPM MSU for interest and valuable discussions. The work was supported in part by RFBR, Grant No. 11-02-00112-a  and by the Russian Ministry of Science and Education, Contract No. 02.740.11.0243.
\end{acknowledgments}


\begin{thebibliography}{200}
\bibitem{Jask}  W.Jaskolski. Phys.Rep. 271 (1996) 1.
\bibitem{Dolmatov}  V.K.Dolmatov, A.S.Baltenkov, J.-P. Connerade, and S.Manson. Rad.Phys. $\&$ Chem. 70 (2004) 417.
\bibitem{AdvQuantChem} J.R. Sabin, E.J. Brandas (eds). Theory of Confined Quantum Systems.   Adv. Quant. Chem., vols. 57-58. Elsevier, Amsterdam, 2009.
\bibitem{Connerade1} J.-P. Connerade, V.K.Dolmatov, P.A.Lakshmi, and S.Manson. J.Phys. B: At. Mol. Opt. Phys. 32 (1999) L239.
\bibitem{Sako} T.Sako and G.H.F. Diercksen. J.Phys. B: At. Mol. Opt. Phys. 36 (2003) 1433; ibid. 36 (2003) 1681.
\bibitem{Maris} H.Maris. Journ. Phys. Soc. Japan. 77 (2008) 80700.
\bibitem{Michels} A.Michels, J. de Boer, and A.Bijl.  Physica (Amsterdam). 4 (1937) 981.
\bibitem{Sommerfeld} A.Sommerfeld, H.Welker. Ann.Phys. 424 (1938) 56.
\bibitem{Aquino} N.Aquino. Adv. Quant. Chem. 57 (2009) 123.
\bibitem{Sen1} H.E. Montgomery, K.D. Sen. Phys. Lett. A 376 (2012) 1992.
\bibitem{WS} E. Wigner, F. Seitz. Phys. Rev. 43 (1933) 804; ibid. 46 (1934) 509.
\bibitem{Alefeld} G. Alefeld and J.Voelkl (eds). Hydrogen in Metals I and II. Springer series Topics in Applied Physics. V. 28-29. Springer, Berlin,  1978.
\bibitem{Fukai} Y.Fukai. The Metal-Hydrogen System, Basic Bulk Properties. Springer, Berlin,  1993.
\bibitem{PdHx} R. Caputo and A. Alavi. Mol. Phys. 101 (2003) 1781.
\bibitem{Wiese} M.H. Al-Hashimi and U.-J.Wiese. ArXiv: 1204.3434v1 [quant-ph].
\bibitem{Sen2} K.D. Sen, V.I. Pupyshev, and H.E. Montgomery. Adv. Quant. Chem. 57 (2009) 25.
\bibitem{LL} L.D.Landau, E.M.Lifshits. Theoretical Physics. v.3. Quantum mechanics. Pergamon, NY, 1974.
\bibitem{MIT bag} A.Chodos, R.L.Jaffe, K.Johnson, C.B.Thorn, and V.F.Weisskopf.  Phys.Rev. D 9 (1974) 3471.
\bibitem{Chiral bag} A.Hosaka, H.Toki. Phys.Rep. 277 (1996) 65; Quarks, baryons and chiral symmetry. World Scientific, Singapoure, 2001.
\bibitem{Pupyshev} V.I. Pupyshev. Rus. J. Phys. Chem. 74 (2000) 50. (Engl. transl.)
\bibitem{PEPAN} K.Sveshnikov, A.Roenko. To be published in Phys.Part.Nucl.Lett. (2013).
\bibitem{Wiese1} M. H. Al-Hashimi and U.-J. Wiese, Ann. Phys. 327 (2012) 1.
\bibitem{Beytmen} H.Beytmen and A.Erdelyi. Higher transcendental functions, v.1. Mc Grow-Hill, NY, 1953.
\bibitem{NW} J. von Neumann and E.P. Wigner. Phys. Z. 30 (1929) 465; ibid. 30 (1929) 467.
\bibitem{Sv-Tol} K.Sveshnikov, A.Tolokonnikov. In preparation.
\end{thebibliography}
\end{document}